\documentclass[aps,prb,amsmath,amssymb,showpacs,reprint]{revtex4-2}
\usepackage{hyperref}
\hypersetup{
pdfstartview=FitH
}
\usepackage{tikz}
\usetikzlibrary{shapes,calc}

\newcommand{\Integer}{\mathbb{Z}}
\DeclareMathOperator{\qdim}{\text{dim}_q}

\DeclareMathOperator{\tr}{{\text{tr}}}

\newcommand{\qsu}{\cU_q\bigl[su(2)\bigr]}
\newcommand{\qg}{\cU_q[\mathfrak{g}]}

\newcommand{\cI}{\mathcal{I}}
\newcommand{\cO}{\mathcal{O}}
\newcommand{\cT}{\mathcal{T}}
\newcommand{\cU}{\mathcal{U}}
\newcommand{\g}{\mathfrak{g}}

\newcommand{\qHaldane}{$q\text{-Haldane}$ }
\newcommand{\qAKLT}{$q\text{AKLT}$ }
\newcommand\qSPT{$q$-SPT }

\begin{document}

\title{Symmetry protected topological phases beyond groups:\\
  The $q$-deformed bilinear-biquadratic spin chain}

\author{Thomas Quella}%
\email{Thomas.Quella@unimelb.edu.au}
\affiliation{The University of Melbourne, School of Mathematics and Statistics, Parkville 3010 VIC, Australia }
 
\date{\today}

\begin{abstract}
  We study the phase diagram of the $SO_q(3)$ quantum group invariant spin-1 bilinear-biquadratic spin chain for real values of $q>1$. Numerical computations suggest that the chain has at least three clearly distinguished phases: A chiral analogue of the Haldane phase, a dimerized phase and a ferromagnetic phase. In contrast, the counterpart of the extended critical region that is known to exist for $q=1$ remains elusive. Our results show that the Haldane phase fails to exhibit a two-fold degeneracy in the entanglement spectrum but that the degeneracy is restored upon a suitable $q$-deformation of the entanglement Hamiltonian which can be interpreted as a Zeeman field. The structure of the phase diagram is confirmed through analytical calculations in the extreme anisotropic limit $q\to\infty$. Our results suggest that symmetries of the form $\qsu$ for distinct choices of $q$ should be interpreted as one single family instead of separate symmetries when defining SPT phases, leading naturally to the notion of a \qSPT phase.
\end{abstract}


\preprint{\eprint{arXiv:yymm.nnnn}}

\maketitle

\section{Introduction}

  According to an old paradigm of Landau's, quantum phases of matter can be classified in terms of the symmetries of the Hamiltonian and the pattern of spontaneous symmetry breaking in the associated manifold of ground states. More recently, it was recognized that, even in the absence of symmetries, quantum systems may exhibit long-range entanglement which allows for a definition of distinct phases which are characterized by specific types of intrinsic topological order \cite{Chen:2010PhRvB..82o5138C}. In addition, symmetries may conspire with topological features, thereby leading to the notions of symmetry protected topological (SPT) phases \cite{Gu:2009PhRvB..80o5131G,Pollmann:2012PhRvB..85g5125P,Schuch:1010.3732v3,Chen:PhysRevB.84.235128,Duivenvoorden:2012arXiv1206.2462D} and symmetry enriched topological (SET) phases \cite{Mesaros:2013PhRvB..87o5115M}. While the former are only short-range entangled and topologically trivial when disregarding symmetries, the latter still may exhibit non-trivial topological order. In all these cases the relevant symmetries are described by groups.

  In a recent paper it has been suggested that there also exist SPT phases that are not protected by group symmetries but rather solely by generalized symmetries such as quantum groups or duality symmetries \cite{Quella:2020PhRvB.102h1120Q}. This claim was verified analytically in the example of an anisotropic deformation of the famous spin-1 AKLT model \cite{Affleck:PhysRevLett.59.799,Affleck:1987cy} whose Hamiltonian is invariant under an action of the quantum group $\qsu$ \cite{Batchelor:1990JPhA...23L.141B,Klumper:1991JPhA...24L.955K,Klumper:1992ZPhyB..87..281K,Batchelor:1994IJMPB...8.3645B,Totsuka:1994JPhA...27.6443T,Fannes:10.1007/BF02101525}. In the current paper we provide additional evidence for this assertion by considering an anisotropic deformation of the general spin-1 bilinear-biquadratic spin chain that still respects the same quantum group symmetry \cite{Batchelor:1990JPhA...23L.141B}. Since these systems are neither exactly solvable nor frustration-free we now rely on suitable numerical methods, specifically iDMRG \cite{Vidal:2007PhRvL..98g0201V,Orus:PhysRevB.78.155117,McCulloch:2008arXiv0804.2509M}, in conjunction with a diagnostic entanglement tool that has been developed in Ref.~\cite{Quella:2020PhRvB.102h1120Q} to support this claim.
  
  Our numerical results show that the $q$-deformed bilinear-biquadratic spin-1 chains exhibits a large chiral analogue of the Haldane phase with a unique and gapped ground state, non-trivial string order \cite{DenNijs:PhysRevB.40.4709,Totsuka:1994JPhA...27.6443T} as well as a two-fold degeneracy in a suitably deformed notion of entanglement spectrum \cite{Quella:2020PhRvB.102h1120Q}. In addition, there is a gapped dimerized phase where the two-fold degeneracy in the entanglement spectrum, both conventional and deformed, is absent and translation symmetry is broken. We interpret this observation as a confirmation that quantum group invariant spin-1 chains admit non-trivial SPT phases, even in the absence of (relevant) ordinary symmetries and even when the ground state features arbitrarily low entanglement. Finally, there is also a gapless ferromagnetic phase. In contrast to the undeformed case $q=1$\cite{Lauchli:2006PhRvB..74n4426L,Manmana:2011PhRvB..83r4433M} the extended critical region between the Haldane phase and the ferromagnetic phase appears to be significantly diminished for $q>1$.

  The basic structure of the phase diagram is confirmed analytically by considering the crystal limit $q\to\infty$ where the Hamiltonian of the $q$-deformed bilinear-biquadratic spin chain becomes trivially diagonal and all ground states as well as excited states can be constructed explicitly.

\section{\label{eq:qGeneral}The $\mathbf{q}$-deformed bilinear-biquadratic spin-1 quantum spin chain}

  In this paper we are interested in the ground state properties of the general XXZ-type anisotropic bilinear-biquadratic spin-$1$ chain with nearest-neighbor interactions (only) that is invariant under the quantum group $\qsu$ as well as translations. For later convenience we will assume that the deformation parameter is expressed as $q=e^\lambda>0$, where $\lambda$ is a real number. With this convention the corresponding Hamiltonian can be written as~\cite{Batchelor:1990JPhA...23L.141B}
\begin{widetext}
\begin{align}
  \label{eq:DeformedHexplicit}
  H^{(\lambda)}
  &=\sum_k\biggl\{
      a\vec{S}_k\cdot\vec{S}_{k+1}
     +b\bigl(\vec{S}_k\cdot\vec{S}_{k+1}\bigr)^2\nonumber\\
   &\qquad\qquad+(\sinh^2\lambda)\Bigl[2a\Bigl(\bigl(S_k^z\bigr)^2+\bigl(S_{k+1}^z\bigr)^2\Bigr)
     +(a-b)\Bigl(S_k^zS_{k+1}^z-\bigl(S_k^z\bigr)^2\bigl(S_{k+1}^z\bigr)^2\Bigr)\Bigr]\nonumber\\
   &\qquad\qquad+\frac{1}{2}(a+b)(\sinh\lambda)\Bigl[\bigl(S_k^xS_{k+1}^x+S_k^yS_{k+1}^y\bigr)\bigl(S_{k+1}^z-S_k^z\bigr)+\text{h.c.}\Bigr]\\
   &\qquad\qquad+2(b-a)(\sinh^2\frac{\lambda}{2})\Bigl[\bigl(S_k^xS_{k+1}^x+S_k^yS_{k+1}^y\bigr)S_k^zS_{k+1}^z+\text{h.c.}\Bigr]\nonumber\\
   &\qquad\qquad+\Bigl(\sinh2\lambda\Bigr)\Bigl[a\bigl(S_{k+1}^z-S_k^z\bigr)+\frac{1}{2}(a+b)S_k^zS_{k+1}^z\bigl(S_{k+1}^z-S_k^z\bigr)\Bigr]
    \biggr\}\;.\nonumber
\end{align}
\end{widetext}
  In what follows we fix the overall energy scale and set $(a,b)=(\cos\theta,\sin\theta)$ such that the angle $\theta$ is the only physically significant parameter of the model. The models with deformation parameters $\lambda$ and $-\lambda$ are physically equivalent (up to inversion) and hence we may assume $\lambda\geq0$. The value of $\lambda$ characterizes the degree of anisotropy. While $\lambda=0$ corresponds to an isotropic chain, the limit $\lambda\to\infty$ is usually referred to as the `extreme anisotropic limit' in related contexts.
  
  The family of Hamiltonians \eqref{eq:DeformedHexplicit} has been studied comprehensively in the past, both for all values of the parameter $\theta$ when $\lambda=0$ but also for $\lambda\neq0$ and (mostly) special values of $\theta$. Focusing on $\lambda\neq0$ for a moment, there exist analytical results on the integrable XXZ spin-1 chain at $\theta=-\frac{\pi}{4}$ \cite{Zamolodchikov:1980ku,Sogo:1984PhLA..104...51S,Bougourzi:1994NuPhB.417..439B,Weston:2006JSMTE..03L.002W}, the \qAKLT model at $\cot\theta=(4\cosh^2\lambda-1)$ \cite{Klumper:1991JPhA...24L.955K,Klumper:1992ZPhyB..87..281K,Klumper:1993EL.....24..293K,Batchelor:1994IJMPB...8.3645B,Totsuka:1994JPhA...27.6443T,Fannes:10.1007/BF02101525,Santos:2012EL.....9837005S,Quella:2020PhRvB.102h1120Q} and a $q$-analogue of the purely biquadratic chain at $\theta=-\frac{\pi}{2}$ \cite{Barber:1989PhRvB..40.4621B,Batchelor:1990PThPS.102...39B}. Some of these results will be reviewed below in Sections~\ref{sc:qAKLT} and~\ref{sc:KnownPhases}.
  
  In the undeformed case $\lambda=0$ we simply recover the standard isotropic bilinear-biquadratic spin chain with Hamiltonian
\begin{align}
  \label{eq:BLBQ}
  H^{(0)}
  &=\sum_k\Bigl\{\cos\theta\,\vec{S}_k\cdot\vec{S}_{k+1}
     +\sin\theta\,\bigl(\vec{S}_k\cdot\vec{S}_{k+1}\bigr)^2\Bigr\}\;.
\end{align}
  It is well-established that this model has a rich phase diagram with various types of gapped and gapless phases, exotic orders, enhanced symmetries and critical points \cite{Affleck:1986NuPhB.265..409A,Lauchli:2006PhRvB..74n4426L,Manmana:2011PhRvB..83r4433M}. Of particular theoretical interest is the Haldane phase in the range $\theta\in(-\frac{\pi}{4},\frac{\pi}{4})$ which is widely regarded as a prototypical example of an SPT phase \cite{Pollmann:2012PhRvB..85g5125P,Pollmann:PhysRevB.81.064439}. Analytical results exist at four integrable points at $\theta\in\bigl\{\pm\frac{\pi}{4},\pm\frac{3\pi}{4}\bigr\}$ \cite{Uimin:1970JETPL..12..225U,Lai:1974JMP....15.1675L,Sutherland:1975PhRvB..12.3795S,Takhtajan:1982PhLA...87..479T,Babujian:1982PhLA...90..479B}, at the AKLT point with $\cot\theta=3$ \cite{Affleck:PhysRevLett.59.799,Affleck:1987cy} and the purely biquadratic spin chain where $\theta=-\frac{\pi}{4}$ \cite{Barber:1989PhRvB..40.4621B,Klumper:1989EL......9..815K,Xian:1993PhLA..183..437X}.
  
  Our analytical and numerical investigations to be reported in Sections~\ref{sc:Limit} and~\ref{sc:Numerics} show that the phase diagram of the bilinear-biquadratic spin chain in large parts remains unaltered if we allow $\lambda$ to be non-zero, the only exception being the extended critical phase. In particular, there is still a phase that displays all defining properties of the Haldane phase. However, as we will point out in Section~\ref{sc:Symmetries} all the symmetries known to protect the Haldane phase are broken and hence we refer to this phase as the \qHaldane phase. It should be noted that inversion and time-reversal symmetries are generically broken by the Hamiltonian~\eqref{eq:DeformedHexplicit}, so the $q$-Haldane phase is, in fact, a chiral phase.

  It is instructive to also consider the limit $\lambda\to\infty$ which leads to drastic simplifications of the $q$-deformed bilinear-biquadratic spin chain. Indeed, after an appropriate rescaling of the Hamiltonians $H^{(\lambda)}$ of Eq.~\eqref{eq:DeformedHexplicit} by a factor $e^{-2\lambda}$ we find
\begin{align}
  \label{eq:LimitHamiltonian}
  H^{(\infty)}
  &=\cos\theta\bigl(S_L^z-S_1^z\bigr)+\sum_k\biggl\{
   4\cos\theta\bigl(S_k^z\bigr)^2\nonumber\\
  &\quad-\sqrt{2}\sin(\theta-\tfrac{\pi}{4})\Bigl(S_k^zS_{k+1}^z-\bigl(S_k^z\bigr)^2\bigl(S_{k+1}^z\bigr)^2\Bigr)\nonumber\\
   &\quad+\frac{1}{\sqrt{2}}\sin(\theta+\tfrac{\pi}{4})S_k^zS_{k+1}^z\bigl(S_{k+1}^z-S_k^z\bigr)
    \biggr\}
\end{align}
  for a finite chain of length $L$. The first term, which corresponds to a boundary magnetic field, vanishes if we impose periodic boundary conditions. Obviously this Hamiltonian is trivially diagonal in the standard spin basis and there is clearly no entanglement in its ground state, regardless of the boundary conditions. We would like to stress that the limit $q\to\infty$ (which is equivalent to $q\to0$) also plays a fundamental role in the theory of quantum groups where it is linked to the theory of `crystal bases' \cite{Kashiwara:1990CMaPh.133..249K,Lusztig:MR1035415} and the associated combinatorial description of representations.
  
  Let us finally mention a few important points of clarification concerning quantum group invariant spin chains. First of all, the quantum group invariance of the Hamiltonian~\eqref{eq:DeformedHexplicit} is entirely unrelated to any form of integrability. In fact, while the definition and solution of quantum integrable models frequently makes use of quantum groups the Hamiltonian in these models is usually {\em not} commuting with the quantum group, so the quantum group is generally {\em not} a symmetry (at least for a finite chain), see however Ref.~\onlinecite{Nepomechie:2018NuPhB.930...91N} and references therein. Secondly, it is well known that quantum group symmetry is not consistent with periodic boundary conditions, except if a suitable (non-local) twist is introduced, see Ref.~\cite{Karowski:1994NuPhB.419..567K,Grosse:1994JPhA...27.4761G,Quella:2020PhRvB.102h1120Q} for details and references. In what follows we will thus only be concerned with infinite chains.

\section{\label{sc:Limit}The phase diagram in the extreme anisotropic limit ${\mathbf\lambda\to\infty}$}

\begin{table}
\begin{center}
\begin{tabular}{c|ccc}
  $(S_1^z,S_2^z)$ & $1$ & $0$ & $-1$ \\\hline
  $1$ & $4\cos\theta$ & $2\cos\theta$ & $3\sqrt{2}\sin(\theta+\tfrac{\pi}{4})$ \\
  $0$ & $2\cos\theta$ & $0$ & $2\cos\theta$\\
  $-1$ & $\sqrt{2}\sin(\theta+\tfrac{\pi}{4})$ & $2\cos\theta$ & $4\cos\theta$
\end{tabular}
  \caption{\label{tb:LimitHamiltonian} The two-site energies in the extreme anisotropic limit $\lambda\to\infty$, see Eq.~\eqref{eq:LimitHamiltonian}. The single-site term $\cos\theta(S_2^z-S_1^z)$ has been dropped since for longer chains it only affects the boundary spins and drops out for periodic boundary conditions.}
\end{center}
\end{table}

  Before diving into numerical investigations we make an effort to obtain some basic analytical understanding of the phase diagram by exploring the extreme anisotropic limit $\lambda\to\infty$. In this limit it is straightforward to determine the ground states for arbitrary values of the angle $\theta$ since the Hamiltonian~\ref{eq:LimitHamiltonian} is diagonal in the standard spin basis. Rather surprisingly, the resulting phase diagram seems to reflect in great detail what will later also be seen for finite values of the anisotropy parameter $\lambda\neq0$.
  
\begin{figure}
\begin{center}
  \includegraphics{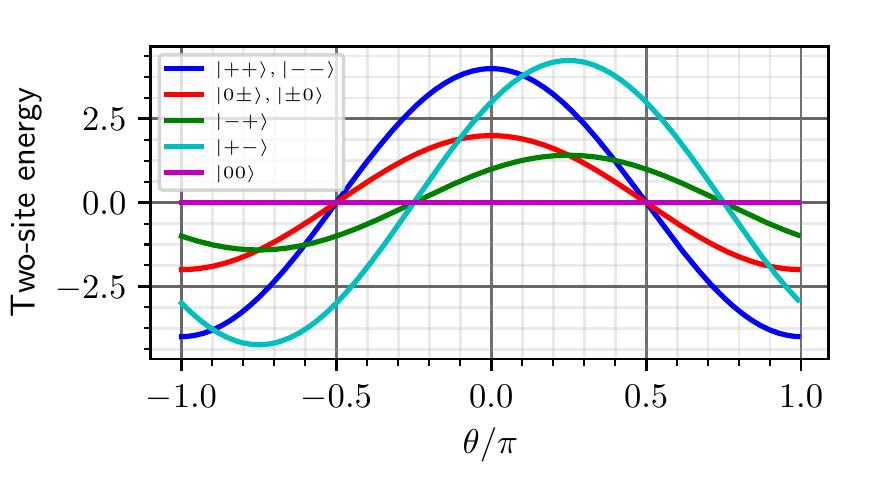}
  \caption{\label{fig:TwoSiteEnergies}(Color online) Plot of the two-site energies for the limiting Hamiltonian $H^{(\infty)}$ as summarized in Table~\ref{tb:LimitHamiltonian}.}
\end{center}
\end{figure}

  The possible energies of the two-site Hamiltonian are summarized in Table~\ref{tb:LimitHamiltonian} and plotted in Figure~\ref{fig:TwoSiteEnergies}. We recognize that the `phase diagram' exhibits three different regions with qualitatively distinct types of ground states. First of all there is a phase with a unique ground state $|00\rangle$ with bond energy $\epsilon=0$ for $\theta\in(-\frac{\pi}{4},\frac{\pi}{4})$. In the interval $(-\xi_c,-\frac{\pi}{4})$ where $\xi_c=-2\arctan(3+\sqrt{10})\approx-0.90\,\pi\approx-2.82$ there is a unique ground state $|{+}{-}\rangle$ with bond energy $\epsilon=3\sqrt{2}\sin(\theta+\frac{\pi}{4})$. (We recall that inversion symmetry is broken so that $|{+}{-}\rangle$ and $|{-}{+}\rangle$ appear on a different footing.) For the remaining parameters the chain is in a ferromagnetic phase with two degenerate Ising-like ground states $|{\pm}{\pm}\rangle$ with bond energy $\epsilon=4\cos\theta$. The value of $\xi_c$ will not play any role in what follows since the location of the transition is shifted to $\theta_c=-\frac{3\pi}{4}$ as the number of sites increases beyond two.

  Let us now turn our attention to the full chain with a length $L>2$. For simplicity we consider periodic boundary conditions first in order to avoid boundary effects resulting from the single-site terms.\footnote{We would like to stress that this means honest periodic boundary conditions, not the non-local twisted boundary conditions that would be appropriate to preserve the quantum group symmetry.} In the limit $\lambda\to\infty$, the total energy is a sum over two-site energies. Hence the previous analysis can simply be extended to the full chain whenever the relevant two-site ground state allows for a frustration-free continuation to the full chain. This is clearly the case for the state $|\cdots000\cdots\rangle$ with bond energy $\epsilon=0$ which hence is the unique ground state in the interval $\theta\in(-\frac{\pi}{4},\frac{\pi}{2})$. Moreover, this is also the case for the two degenerate ferromagnetic Ising-like ground states $|\cdots{\pm}{\pm}{\pm}\cdots\rangle$ with bond energy $\epsilon=4\cos\theta$.

\begin{figure}
\begin{center}
  \includegraphics{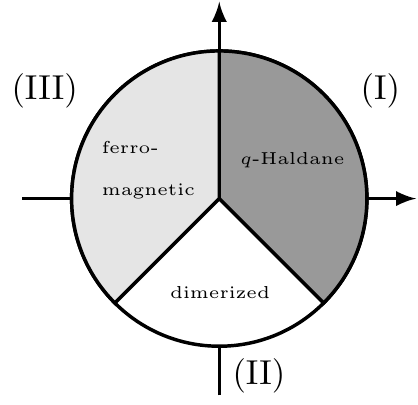}
\caption{\label{fig:PhaseDiagramCrystalLimit}(Color online) The phase diagram of the $q$-deformed bilinear-biquadratic spin chain in the crystal limit $\lambda\to\infty$. The ground states in the three different regions are (I) $|\cdots0000\cdots\rangle$, (II) $|\cdots{\pm}{\mp}{\pm}{\mp}\cdots\rangle$ and (III) $|\cdots{\pm}{\pm}{\pm}{\pm}\cdots\rangle$.}
\end{center}
\end{figure}

  However, it is obviously impossible to extend the state $|{+}{-}\rangle$ to the full chain without obtaining frustrated contributions from $|{-}{+}\rangle$. As a consequence of this unavoidable frustration the phase boundaries between this phase and the other two phases are shifted. It turns out that, for an even number of sites, the appropriate ground states are $|\cdots{\pm}{\mp}{\pm}\cdots\rangle$ with average (frustrated) bond energy
\begin{align}
  \epsilon
  =2\sqrt{2}\sin(\theta+\tfrac{\pi}{4})
  =2(\sin\theta+\cos\theta)\;.
\end{align}
  Obviously, both of these states break translation invariance and the new phase boundary can easily be seen to be located at $\theta_c=-\frac{3\pi}{4}$. For an odd number of sites the actual ground states show a slightly higher degree of degeneracy, frustration and hence energy, and therefore the corresponding states should not be regarded as ground states in the thermodynamic limit. Our findings about the phase diagram are depicted in Figure~\ref{fig:PhaseDiagramCrystalLimit}.
   
  Besides looking at the invididual phases it is also instructive to understand what happens at the phase boundaries. The ground state bond energies derived above exhibit crossings and this means that the phase transitions are first order transitions. This is not surprising since the Hamiltonian is trivially diagonal and the ground state is constant within each phase, so there is no continuous transition. In view of the required rescaling of the Hamiltonian when performing the limit $\lambda\to\infty$ and the known -- but different -- results at $\lambda=0$ it is questionable though whether this should be interpreted as an indication about the nature of the phase transitions also for finite values of $\lambda$. It is also worth mentioning that the ground state degeneracies are drastically enlarged at the two points $\theta\in\{-\frac{3\pi}{4},\frac{\pi}{2}\}$.
  
  Let us finally briefly comment on the situation with open boundary conditions. In that case, the analysis essentially parallels our previous discussion except for the presence of boundary terms whose effect on the energy becomes neglegible in the thermodynamic limit $L\to\infty$. However, for small values of $L$ explicit calculation of the corresponding energies shows that the transition between the `dimerized phase' (whose degeneracy is actually lifted by the boundary terms)
  and the `Haldane phase' shifts from $-\frac{\pi}{4}$ to slightly larger values. We are tempted to speculate about a relation between this observation and earlier claims regarding the existence of a potential additional phase between the ferromagnetic and the dimerized phase in the undeformed bilinear-biquadratic spin chain.\cite{Lauchli:2006PhRvB..74n4426L} Our analysis here was concerned with the limit $q\to\infty$ and not with the limit $q\to1$. However, if it has any implications for the aforementioned question our discussion of periodic boundary conditions would suggest that there should be no such intermediate phase. We also note the absence of a separate phase in the region $\theta\in(\frac{\pi}{4},\frac{\pi}{2})$ which is known to exist for $\lambda=0$, see e.g.\ Refs.~\cite{Lauchli:2006PhRvB..74n4426L,Manmana:2011PhRvB..83r4433M}.

  The crystal limit $\lambda\to\infty$ has been successfully used in the past to obtain a combinatorial description of eigenstates of integrable Hamiltonians ~\cite{Kang:1992IJMPA...7S.449K,Davies:1993CMaPh.151...89D,Nakayashiki:1996CMaPh.178..179N,Hong:1998math.....11175H,Lamers:2020arXiv200413210L}. For the purposes of the present paper we went a slightly different route and focused on the ground state(s) only. However, in contrast to the papers cited above we did not restrict our attention to a single point in the phase diagram but rather used the information about the ground states to map the latter out in full detail. It can be expected that the information we obtained here for $\lambda\to\infty$ reflects what is happening at finite values of $\lambda$ and our numerical analysis largely confirms that this is indeed the case, see Section~\ref{sc:Numerics}.

\section{\label{sc:Symmetries}Symmetries of the $\mathbf{q}$-deformed bilinear-biquadratic spin-1 chain}

  The ordinary isotropic bilinear-biquadratic spin chain associated with $\lambda=0$ has a number of symmetries that are crucial in explaining its physical properties. It first of all exhibits an $SO(3)$ spin-rotation and a $\Integer_2$ spin-flip symmetry which can be combined into an $O(3)$ symmetry. Moreover, it has a number of space-time symmetries, specifically invariance under translations by one site, time-reversal $\cT$ and inversion $\cI$. It is well-established that the Haldane phase of this chain can be regarded as an SPT phase with respect to either of the following symmetries \cite{Pollmann:2012PhRvB..85g5125P}: Spin-rotations $SO(3)$, its $\Integer_2\times\Integer_2$ subgroup of $\pi$-rotations around the principal axes, time-reversal and inversion. The key to this insight is the presence of non-trivial projective representations of these symmetry groups. It has also been established that the key signatures of that phase are a two-fold degeneracy in the bipartite ground state entanglement spectrum \cite{Pollmann:PhysRevB.81.064439} and the existence of quantized topological invariants \cite{Pollmann:2012PhRvB..86l5441P}.\footnote{As emphasized in \cite{Pollmann:PhysRevB.81.064439}, other signatures such as non-local string order and fractionalized boundary spins may well be absent under deformations that break some (but not all) of these symmetries.}

  In contrast, for generic values of $\lambda$ most of the symmetries mentioned in the previous paragraph are explicitly broken by the Hamiltonian~\eqref{eq:DeformedHexplicit}. In view of the anisotropy the spin-rotation symmetry is broken to a $U(1)$ subgroup of rotations around the $z$-axis which also contains one factor $\Integer_2$ of the dihedral group $\Integer_2\times\Integer_2$. However, the second factor $\Integer_2$ as well as spin-flip, time-reversal and inversion symmetry are broken by term such as $S_k^zS_{k+1}^z\bigl(S_{k+1}^z-S_k^z\bigr)$.\footnote{We note that some of these symmetries are restored for $\theta=-\frac{\pi}{4}$ and $\theta=\frac{3\pi}{4}$.}
  
  We would like to emphasize that neither of the remaining symmetries usually considered allows for non-trivial projective representations. In other words: All the standard symmetries that are known to protect the topological properties of the Haldane phase \cite{Gu:2009PhRvB..80o5131G,Pollmann:2012PhRvB..85g5125P} cease to exist. According to the classification of Refs.~\onlinecite{Schuch:1010.3732v3,Chen:PhysRevB.84.235128,Duivenvoorden:2012arXiv1206.2462D} there is hence no reason to think of any of the Hamiltonians $H^{(\lambda)}$ as realizing an SPT phase whenever $\lambda\neq0$.
  
  However, by construction the Hamiltonian~\eqref{eq:DeformedHexplicit} is invariant under the $q$-deformed symmetry $\qsu$ and it has recently been argued that this generalized symmetry is (for chains of integer physical spins) still capable of protecting the relevant topological properties \cite{Quella:2020PhRvB.102h1120Q}. We also note that the breaking of the usual discrete symmetries is relatively mild. Indeed, all of them can be restored when combining them with a duality transformation $\lambda\to-\lambda$ that changes the coupling. It was suggested in Ref.~\cite{Quella:2020PhRvB.102h1120Q} that these duality-type symmetries by themselves might already be sufficient to guarantee the protection of non-trivial topological phases.

\section{\label{sc:OrderParameters}Order parameters and $\mathbf{q}$-deformed entanglement spectrum}

\begin{figure}[t]
\begin{center}
  \includegraphics{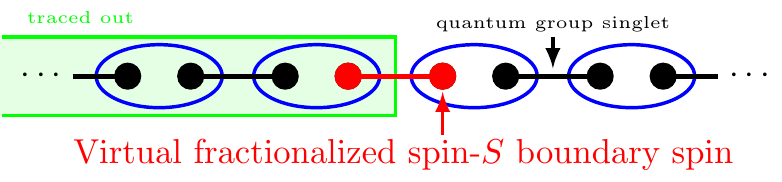}
  \caption{\label{fig:qAKLT}(Color online) Bipartite entanglement cut in the spin-2S \qAKLT state on an infinite chain.}
\end{center}
\end{figure}

  The phase diagram of the usual bilinear-biquadratic spin chain can be explored using a number of diagnostic tools, including order parameters to detect spontaneous symmetry breaking and entanglement measures. In what follows we will think of all these quantities as being calculated from an iMPS representation\cite{Vidal:2007PhRvL..98g0201V,Orus:PhysRevB.78.155117} of its ground state. First of all, the entanglement entropy or rather its (non-)scaling with the bond dimension can be used to determine whether the system is gapped or critical and, in the latter case, to infer the value of the central charge \cite{Tagliacozzo:2008PhRvB..78b4410T,Pollmann:2009PhRvL.102y5701P}. A closely related indicator is the behavior of the correlation length. Both quantities can easily be computed from the iMPS representation and the eigenvalues of the associated transfer matrix. Additional methods for the analysis of critical systems have been described in Refs.~\cite{Pollmann:2010NJPh...12b5006P,Pirvu:2012PhRvB..86g5117P,Stojevic:2015PhRvB..91c5120S}.
  
  For an infinite critical system with central charge $c$ one has the scaling relations~\cite{Korepin:PhysRevLett.92.096402,Calabrese:2004JSMTE..06..002C,Tagliacozzo:2008PhRvB..78b4410T,Pollmann:2009PhRvL.102y5701P}
\begin{align}
  \label{eq:EntanglementScaling}
  S(\chi)\sim\frac{c\kappa}{6}\log\chi\;,\qquad
  \xi(\chi)\sim\xi_0\chi^\kappa\;.
\end{align}
  Here $S(\chi)$ is the entanglement entropy, $\xi(\chi)$ is the correlation length and $\chi$ is the bond dimension used for the iMPS in the simulation of the model. Plotting $S$ and $\log\xi$ as a function of $\log\chi$ one should obtain two straight lines and from the respective slopes one can infer the central charge $c$.
  
  Beyond this, different phases can also be distinguished by means of order parameters such as ferromagnetic order, dimerization, string order \cite{DenNijs:PhysRevB.40.4709} as well as degeneracies in the entanglement spectrum \cite{Pollmann:PhysRevB.81.064439,Quella:2020PhRvB.102h1120Q}. We now introduce these quantities in more detail, with an emphasis on the appropriate $q$-deformed setting. This is particularly important for the entanglement spectrum, where the necessary adjustment is responsible for additional Zeeman-like terms in the entanglement Hamiltonian.

  Ferromagnetic ordering can be detected using the order parameter
\begin{align}
  \label{eq:FMO}
  \cO_{\text{FMO}}^\alpha
  =\lim_{|i-j|\to\infty}\langle S_i^\alpha S_j^\alpha\rangle
  \quad(\text{with }\alpha=x,y,z)\;.
\end{align}
  It should be noted that, in the present case, this order parameter depends on the direction singled out since the $q$-deformed chains are anisotropic and rotation symmetry is broken. Dimerization in singlet ground states is usually measured in terms of the order parameter $\bigl\langle(\vec{S}_i\cdot\vec{S}_{i+1})-(\vec{S}_{i+1}\cdot\vec{S}_{i+2})\bigr\rangle$. However, in view of the fact that rotation symmetry is anyway broken by the anisotropy we will measure breaking of translation symmetry by the simpler order parameter
\begin{align}
  \cO_{\text{DO}}^\alpha
  =\lim_{|i-j|\to\infty}\bigl\langle S_i^\alpha S_{i+1}^\alpha-S_{i+1}^\alpha S_{i+2}^\alpha\bigr\rangle
\end{align}
  instead of adjusting the scalar products $\vec{S}_i\cdot\vec{S}_{i+1}$ appropriately to the $q$-deformed setting. Finally, it is known that the non-local string order parameter
\begin{align}
  \cO_{\text{SO}}
  =\lim_{|i-j|\to\infty}
    \Bigl\langle S_i^z\prod_{k=i+1}^{j-1}e^{\pi i S_k^z}S_j^z\Bigr\rangle\;,
\end{align}
  can be used to measure the diluted antiferromagnetic order present in the Haldane phase of the bilinear-biquadratic spin chain \cite{DenNijs:PhysRevB.40.4709}. This is known to carry over to the \qAKLT model \cite{Totsuka:1994JPhA...27.6443T} and hence this seems like a sensible choice also for other values of $\theta$.

  Let us finally introduce a $q$-deformed notion of the usual bipartite entanglement spectrum that has been shown to lead to a two-fold degeneracy for odd-spin \qAKLT states, while even-spin \qAKLT states do not exhibit that degeneracy \cite{Quella:2020PhRvB.102h1120Q}. We will later confirm that this two-fold degeneracy is not only present at the \qAKLT point of the $q$-deformed bilinear-biquadratic spin chain but everywhere in the \qHaldane phase and absent in the dimerized phase. This provides additional evidence for the claim made in Ref.~\cite{Quella:2020PhRvB.102h1120Q} that it can be interpreted as a signature of non-trivial topology, thereby generalizing earlier findings for the isotropic chain \cite{Pollmann:PhysRevB.81.064439}.

\begin{figure}
\begin{center}
  \includegraphics{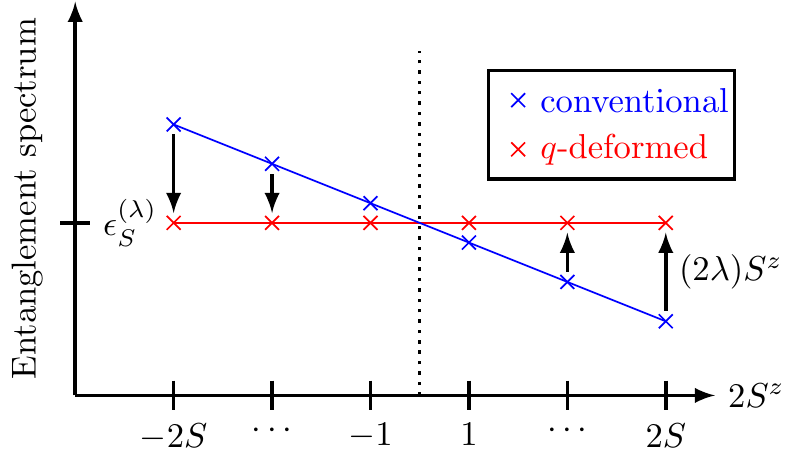}
  \caption{\label{fig:ES_tilt}(Color online) Conventional and deformed symmetry-resolved entanglement spectrum for spin-$2S$ qAKLT states with $q>1$ (where $\epsilon_S^{(\lambda)}=\log[2S+1]_\lambda$). The $q$-deformation clearly acts as a Zeeman field in the entanglement Hamiltonian. A two-fold degeneracy (half-integral $S$) signals non-trivial topology.}
\end{center}
\end{figure}
  
  Let $|\psi\rangle$ be the ground state of the system (assumed to be unique and hence a $\qsu$-singlet) and $\rho=|\psi\rangle\langle\psi|$ be the associated density matrix. The bipartite entanglement in that state can be captured in a $\qsu$-invariant fashion by means of a $q$-deformed generalization of the reduced density matrix. Following Ref.~\cite{Couvreur:2017PhRvL.119d0601C} we define
\begin{align}
  \label{eq:EntanglementSpectrum}
  \rho_R^{(\lambda)}
  =\tr_L\bigl(e^{-2\lambda S_L^z}\rho\bigr)\;,
\end{align}
  where $L$ and $R$ refer to the left and right semi-infinite halfs, repectively, and $S_L^z$ corresponds to the action of $S^z$ on the left part of the chain which is traced out. The eigenvalues of $\rho_R^{(\lambda)}$ define what will be referred to as the $q$-deformed entanglement spectrum and denoted by $\epsilon_\alpha^{(\lambda)}$.
  
  We also define an associated $q$-deformed entanglement entropy  which is given by~\cite{Couvreur:2017PhRvL.119d0601C}
\begin{align}
  \label{eq:EntanglementEntropy}
  S_{EE}^{(\lambda)}
  =-\tr_R\bigl(e^{2\lambda S_R^z}\rho_R^{(\lambda)}\log\rho_R^{(\lambda)}\bigr)\;.
\end{align}
  The $q$-deformed entanglement measures just defined have previously been calculated in Ref.~\cite{Quella:2020PhRvB.102h1120Q} for integer spin \qAKLT states, where $S$ is the auxiliary spin and $2S$ is the integer physical spin, see Figure~\ref{fig:qAKLT} for an illustration. Both quantities are given by the logarithm of the quantum dimension, $\log\qdim(S)$, where $\qdim(S)=[2S+1]_\lambda$ and $[x]_\lambda=\sinh(\lambda x)/\sinh(\lambda)$ denotes a $q$-number. We would like to emphasize that the $q$-deformed entanglement spectrum is fully degenerate even though the conventional entanglement entropy may be arbitrarily small and in fact vanishes in the crystal limit $\lambda\to\infty$.
  
  In anticipation of our numerical experiments let us briefly explain the main difference between the conventional and the $q$-deformed notions of entanglement. We recognize from Eq.~\eqref{eq:EntanglementSpectrum} that the insertion of the factor $e^{-2\lambda S_L^z}$ (which for a singlet can be rewritten as $e^{2\lambda S_R^z}$) may be interpreted as the addition of a uniform magnetic field in the entanglement Hamiltonian that is coupled to the $z$-component of the individual spins. But while such a magnetic field usually, e.g.\ in systems like the hydrogen atom, has the effect of breaking rotation symmetry and destroying the associated degeneracies, here it has precisely the opposite effect of actually restoring degeneracies associated with irreducible representations of $\qsu$. For spin-$2S$ \qAKLT states this is illustrated in Figure~\ref{fig:ES_tilt} and the same effect can also be seen in our numerical experiments, see Figure~\ref{fig:ESall}.  

\section{\label{sc:qAKLT}The $\mathbf{q}$-deformed AKLT model}

  The \qAKLT model corresponds to the point $\cot\theta=(4\cosh^2\lambda-1)$ (with $\theta\in(0,\frac{\pi}{2})$) and has been studied from a variety of perspectives \cite{Klumper:1991JPhA...24L.955K,Klumper:1992ZPhyB..87..281K,Klumper:1993EL.....24..293K,Batchelor:1994IJMPB...8.3645B,Totsuka:1994JPhA...27.6443T,Fannes:10.1007/BF02101525,Santos:2012EL.....9837005S,Quella:2020PhRvB.102h1120Q}. In this section we summarize its most important properties.

  The ground state of the \qAKLT model is described by an MPS (or iMPS \cite{Quella:2020PhRvB.102h1120Q}) with bond dimension $\chi=2$ \cite{Klumper:1991JPhA...24L.955K,Klumper:1992ZPhyB..87..281K,Totsuka:1994JPhA...27.6443T}. From the associated transfer matrix one can infer the correlation length $\xi$ which is known to be given by
\begin{align}
  \xi^{-1}
  =\log[3]_\lambda
  =\log(1+2\cosh\lambda)
  \xrightarrow{\ \lambda\to0\ }\log(3)\;.
\end{align}
  The same correlation length also determines the exponential decay of the two-point functions $\langle S_i^z S_j^z\rangle$ and $\langle S_i^xS_j^x\rangle$ which thus shows the absence of ferromagnetic order, i.e.\ $\cO_{\text{FMP}}^\alpha=0$. Moreover, given that the ground state of the \qAKLT model is translation invariant we clearly have $\cO_{\text{DO}}^\alpha=0$. The degree of anisotropy can be measured by considering the single-site expectation values $\bigl\langle(S^z)^2\bigr\rangle=2/(3+4\sinh^2\lambda)\leq\frac{2}{3}$ and $\bigl\langle(S^x)^2\bigr\rangle=\bigl\langle(S^y)^2\bigr\rangle=1-\frac{1}{2}\bigl\langle(S^z)^2\bigr\rangle\geq\frac{2}{3}$ and these show that the $xy$-plane is favored over the $z$-direction \cite{Klumper:1992ZPhyB..87..281K}. Finally, the string order parameter can be calculated exactly and one finds \cite{Totsuka:1994JPhA...27.6443T}
\begin{align}
  \cO_{\text{SO}}
  =-\biggl(\frac{2}{[3]_\lambda}\biggr)^2
  \xrightarrow{\ \lambda\to0\ }-\frac{4}{9}\;.
\end{align}
  This can be interpreted as the presence of a diluted antiferromagnetic order that is built into the valence-bond description of the \qAKLT state.

  The iMPS also determines the entanglement properties. The $q$-deformed entanglement energies and entropy coincide and read \cite{Quella:2020PhRvB.102h1120Q}
\begin{align}
  S_{EE}^{(q)}
  =\epsilon_\pm^{(q)}
  =\log(2\cosh\lambda)
  \xrightarrow{\ \lambda\to0\ }\log2\;.
\end{align}
  The most important characteristic of this result is the two-fold degeneracy of the $q$-deformed entanglement spectrum. In contrast, the conventional entanglement spectrum is non-degenerate and the associated entanglement entropy~\cite{Quella:2020PhRvB.102h1120Q}
\begin{align}
  S_{EE}
  =\log(1+e^{2\lambda})-\lambda(1+\tanh\lambda)
\end{align}
  shows that the \qAKLT state has arbitrarily low entanglement as $\lambda\to\infty$. This observation is in agreement with our earlier observation that all eigenstates of the Hamiltonian are trivially product states in the extreme anisotropic limit, see Section~\ref{sc:Limit}. We note that the $q$-deformed entanglement spectrum cannot be defined anymore once the limit $\lambda\to\infty$ has been performed. On the other hand, the degeneracy in the $q$-deformed entanglement spectrum is {\em always} present for finite values of $\lambda$, no matter how large. This reflects the fact that the system exhibits entanglement that cannot be removed in a $\qsu$-invariant way \cite{Quella:2020PhRvB.102h1120Q}.

  As a sanity check we have determined the ground state of the \qAKLT Hamiltonian using iDMRG~\cite{Crosswhite:2008PhRvB..78c5116C,McCulloch:2008arXiv0804.2509M,Schollwock:2011AnPhy.326...96S} for selected values of $\lambda$. This calculation confirmed the theoretical predictions for the order parameters as well as the entanglement spectrum and entropy to very high numerical precision (less than $10^{-10}$). These calculations were performed using the TeNPy Library \cite{TeNPy} (version 0.6.1).

\section{\label{sc:KnownPhases}Known results about the phase diagram}

  The phase diagram of the conventional bilinear-biquadratic spin chain ($\lambda=0$ in our conventions) has been studied extensively in the past, see e.g.~\cite{Affleck:1986NuPhB.265..409A,Lauchli:2006PhRvB..74n4426L,Manmana:2011PhRvB..83r4433M} and references therein. In this Section we review some of the known results away from the AKLT point \cite{Affleck:PhysRevLett.59.799,Affleck:1987cy} that has already been covered in Section~\ref{sc:qAKLT}.
  
\begin{figure}
\begin{center}
  \includegraphics{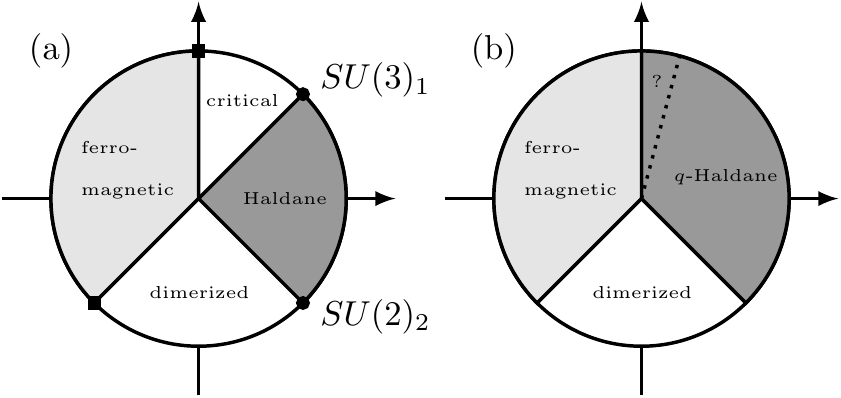}
\caption{\label{fig:PhaseDiagram}(Color online) (a) The known phase diagram for the conventional and (b) and the suggested phase diagram for the $q$-deformed bilinear-biquadratic spin chain. The size of the unascertained region marked with `?' depends on $\lambda$ and vanishes as $\lambda\to\infty$.}
\end{center}
\end{figure}

  The conventional bilinear-biquadratic spin chain has four distinct phases, see Figure~\ref{fig:PhaseDiagram}. The chain is in the gapped Haldane phase with a unique $SO(3)$-invariant ground state for $\theta\in(-\frac{\pi}{4},\frac{\pi}{4})$. Adjacent to this phase are a dimerized phase with spontaneously broken translation symmetry for $\theta\in(-3\frac{\pi}{4},-\frac{\pi}{4})$ and an extended critical phase with spin nematic correlations for $\theta\in(\frac{\pi}{4},\frac{\pi}{2})$. Finally, the zero-temperature chain exhibits a ferromagnetic phase in the rest of the phase diagram.
  
  The phase transitions from the Haldane phase to the neighboring phases are described by well-known integrable models. At $\theta=-\frac{\pi}{4}$ the bilinear-biquadratic spin chain reduces to the Babujian-Takhtajan chain~\cite{Takhtajan:1982PhLA...87..479T,Babujian:1982PhLA...90..479B} which is known to be described by the $SU(2)_2$ WZW conformal field theory with $c=\frac{3}{2}$ in the thermodynamic limit \cite{Affleck:1986bv,Affleck:1987PhRvB..36.5291A}. At $\theta=\frac{\pi}{4}$ the $SU(2)$-symmetry is enhanced to $SU(3)$ and the spin chain reduces to the Uimin-Lai-Sutherland chain \cite{Uimin:1970JETPL..12..225U,Lai:1974JMP....15.1675L,Sutherland:1975PhRvB..12.3795S}, which is known to be described by a $SU(3)_1$ WZW conformal field theory with $c=2$ in the thermodynamic limit. An enhanced $SU(3)$ symmetry is obviously also present at the opposite site for $\theta=-\frac{3\pi}{4}$ where the phase transition from the dimerized phase to the ferromagnetic phase takes place.\footnote{There was some debate in the literature whether there is an additional gapped spin-nematic phase in the interval $(-\frac{3\pi}{4},-\frac{\pi}{2})$, see Ref.~\cite{Lauchli:2006PhRvB..74n4426L}, but since this is not relevant for our paper we do not discuss this in more detail.}
  
  We note that the $SU(3)_1$ WZW model can be interpreted as an $SU(2)_4$ WZW model with a non-diagonal modular invariant. The fact that the levels of the two $SU(2)_2$ and $SU(2)_4$ WZW models describing the phase transitions above are multiples of two and the associated multicriticality are not accidental. Rather, they are a direct consequence of fact that these are topological phase transitions connecting a trivial and a non-trivial topological phase \cite{Roy:2015arXiv151205229R} (see also \cite{Furuya:2017PhRvL.118b1601F,Lecheminant:2015NuPhB.901..510L}).

  Considerable analytical information is also available for the purely biquadratic chain at $\theta=-\frac{\pi}{2}$ which has been related to a 9-state Potts model and to an integrable spin-$\frac{1}{2}$ XXZ model with twisted boundary conditions \cite{Barber:1989PhRvB..40.4621B,Klumper:1989EL......9..815K}. In particular, it has been established that the chain is gapped and dimerized at this point, see also \cite{Xian:1993PhLA..183..437X} for independent arguments.
  
  Shifting our attention to the $q$-deformed anisotropic bilinear-biquadratic chain there exists an integrable deformation of the Babujian-Takhtajan chain which is generally known as the integrable Zamolodchikov-Fateev chain or simply as the spin-1 XXZ chain, see Refs.~\cite{Zamolodchikov:1980ku,Sogo:1984PhLA..104...51S,Pasquier:1990NuPhB.330..523P,Bougourzi:1994NuPhB.417..439B,Vlijm:2014JSMTE..05..009V} and references therein. As for $\lambda=0$ it resides at $\theta=-\frac{\pi}{4}$ and has a unique singlet ground state. However, in contrast to the case $\lambda=0$ it turns out to be gapped for $\lambda\neq0$ \cite{Sogo:1984PhLA..104...51S}. Surprisingly, the scaling of the groundstate's bipartite entanglement entropy still indicates a relation to the $SU(2)_2$ WZW model \cite{Weston:2006JSMTE..03L.002W}. It is worth emphasizing that inversion and time reversal symmetry are restored for $\theta=-\frac{\pi}{4}$ which results in $a+b=\sqrt{2}\sin(\theta+\frac{\pi}{4})=0$ in the Hamiltonian~\eqref{eq:DeformedHexplicit}. An anisotropic analogue of the purely biquadratic spin chain has been discussed in Ref.~\cite{Batchelor:1990PThPS.102...39B} and it was shown that this system has a gap. The vicinity of this point has also been explored numerically \cite{Chen:2020PhRvB.102h5146C}.

  In Section~\ref{sc:Limit} we have solved the model in the extreme anisotropic limit $\lambda\to\infty$ and obtained a phase diagram that matches the one for $\lambda=0$ precisely (including the location of phase transitions) except for the absence of the extended critical phase in the region $\theta\in(\frac{\pi}{4},\frac{\pi}{2})$.
  Since the ground states are known exactly it is straightforward to compute various types of order parameters.
  We wish to stress though that the $q$-deformed entanglement spectrum cannot be recovered once the limit $\lambda\to\infty$ has been taken (see also our discussion in Section~\ref{sc:qAKLT}). Hence the fact that the ground state is a product state in the region $\theta\in(-\frac{\pi}{4},\frac{\pi}{2})$ of the Haldane phase is not in conflict with our findings for finite values of $\lambda$.

\section{\label{sc:Numerics}Numerical results}

\begin{figure*}
\begin{center}
  \includegraphics{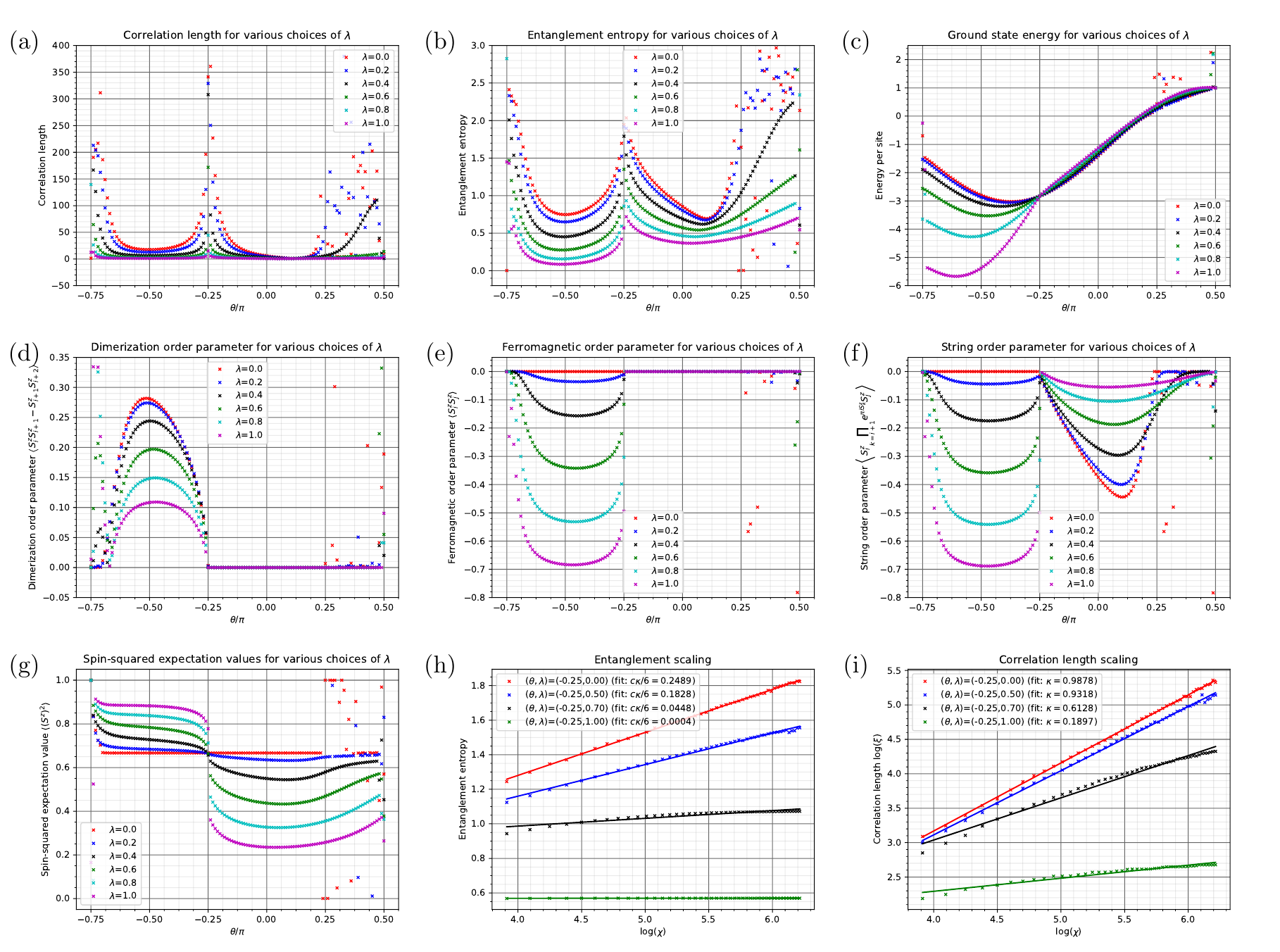}
  \caption{\label{fig:PTall}(Color online) Order parameters and signatures of phase transitions for various values of $\lambda$. For $\lambda=0$ the diagrams show clear evidence of phase transitions at $\theta\in\{-\frac{3\pi}{4},\pm\frac{\pi}{4},\frac{\pi}{2}\}$. For $\lambda>0$, most of these phase transitions clearly persist. However, the phase transition at $\theta=\frac{\pi}{4}$ is significantly shifted to a larger values in a $\lambda$-dependent fashion and becomes very small for large values of $\lambda$. Since convergence was rather bad in the ferromagnetic regime, the corresponding part of the phase diagram is not shown. All simulations with fixed bond dimension used $\chi=800$. Order parameters which involve a limit $|i-j|\to\infty$ are displayed for $|i-j|=1000$.}
\end{center}
\end{figure*}

  To verify the theoretical predictions and plot out part of the phase diagram of the $q$-deformed bilinear-biquadratic spin chains we ran a number of computer simulations. We used iDMRG~\cite{Crosswhite:2008PhRvB..78c5116C,McCulloch:2008arXiv0804.2509M,Schollwock:2011AnPhy.326...96S} to determine the correlation length, expectation values, correlation functions as well as conventional and $q$-deformed entanglement spectra and entropies. All calculations were performed using the TeNPy Library (version 0.6.1) \cite{TeNPy}. This library allows to take into account abelian symmetries such as spin rotations about the $z$-axis that are generated by the component $S_{\text{tot}}^z$ of the total spin.
  
  For our purposes it was particularly important that TeNPy is capable of computing the symmetry-resolved entanglement spectrum, i.e.\ the (conventional) entanglement spectrum organized in terms of the conserved $S^z$-quantum number. For a singlet ground state with $0=S_{\text{tot}}^z=S_L^z+S_R^z$ this permits to post-process the conventional entanglement data using the equations
\begin{align}
  \rho_R^{(\lambda)}
  =e^{2\lambda S_R^z}\rho_R
  \quad\text{ and }\quad
  H_E^{(\lambda)}
  =H_E+2\lambda S_R^z
\end{align}
  that follow directly from Eq.~\eqref{eq:EntanglementSpectrum}. Here $\rho_R$ is the conventional reduced density matrix and $H_E$ ($H_E^{(\lambda)}$) is the ($q$-deformed) entanglement Hamiltonian.
  
  In order to get a basic picture of the phase diagram we determined a ground state for 200 equidistant values of the parameter $\theta\in[-\pi,\pi]$ using iDMRG with bond dimensions $\chi=200$ and then similarly for the smaller interval $\theta\in[-\frac{3\pi}{4},\frac{\pi}{2}]$ with $\chi=800$. Except for the known extended critical region $[\frac{\pi}{4},\frac{\pi}{2}]$, these simulations converged reasonably well in the window $\theta\in(-\frac{3\pi}{4},\frac{\pi}{2})$ and clearly showed phase transitions at $\theta\in\{-\frac{3\pi}{4},-\frac{\pi}{4}\}$, see Figure~\ref{fig:PTall}. In contrast, simulations in the ferromagnetic regime did not converge well, probably due to the fact that there is a highly degenerate ground state manifold and this phase is (trivially) gapless.
  
  The main numerical results are depicted in Figure~\ref{fig:PTall}. Subfigures (a) and (b) show the correlation length and the entanglement entropy which seem to diverge at the points $\theta=-\frac{3\pi}{4}$ and $\theta=-\frac{\pi}{4}$ signaling a quantum phase transition. For $\lambda=0$ and potentially also for small values of $\lambda\gtrsim0$ they also diverge at or in the vicinity of $\theta=\frac{\pi}{4}$ and scattered points indicate convergence issues (here and also in other diagrams). However, for increasing values of $\lambda$ this singular behavior seems to disappear even though some convergence issues remain in the region $\theta\lesssim\frac{\pi}{2}$. We note that the correlation length and the entanglement entropy both decrease as $\lambda$ increases. Our discussion of the crystal limit in Section~\ref{sc:Limit} shows that both tend to zero as $\lambda\to\infty$.

\begin{figure*}
  \includegraphics{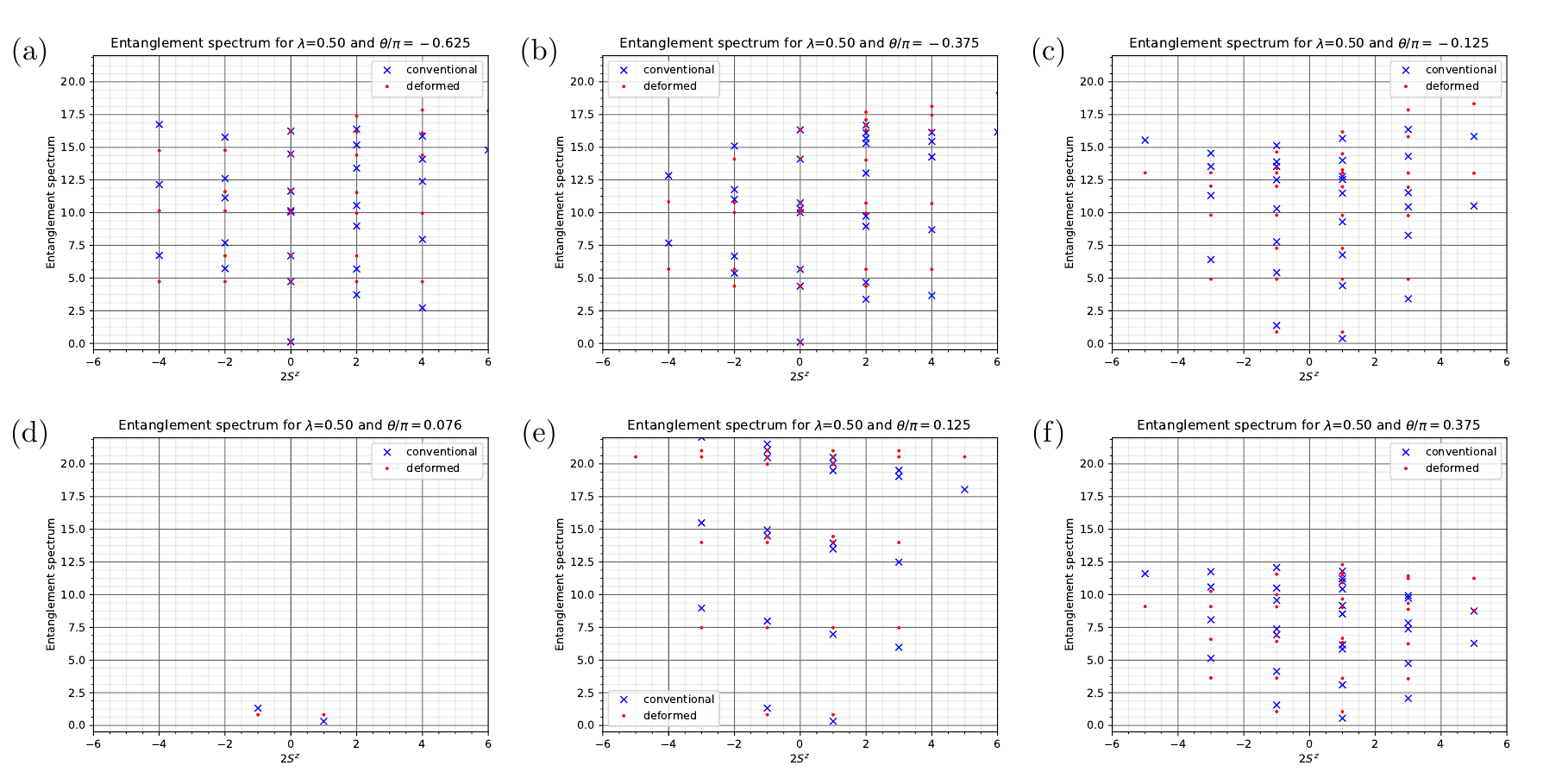}
\begin{center}
  \caption{\label{fig:ESall}(Color online) Symmetry-resolved entanglement spectrum for $\lambda=0.5$ and different values of $\theta$. The degeneracy is absent or odd in the dimerized phase in parts (a) and (b) while it is even in the \qHaldane phase in parts (c)-(f), signalling the presence of non-trivial topology. Part (d) depicts the \qAKLT point at $\theta\approx0.0763979$ and the 2-fold degeneracy associated with its virtual spin-$\frac{1}{2}$ boundary spin.}
\end{center}
\end{figure*}
  
  The ground state energy per site is displayed in subfigure (c). Again we see occasional convergence issues and otherwise a transition from a reasonably smooth dependence on $\theta$ for small values of $\lambda$ to one which seems to have a jump in the first derivative at $\theta=-\frac{\pi}{4}$ for large values of $\lambda$, indicating a first order phase transition, just as predicted by our analysis of the extreme anisotropic limit in Section~\ref{sc:Limit}. It should be noted though that the crystal limit was taken after rescaling the Hamiltonian by a $\lambda$-dependent factor, so the respective energy diagrams are not directly comparable.
  
  The dimerization order parameter is pictured in subfigure (d). It vanishes in the Haldane phase and is non-zero in the dimerized phase, just as expected. The same is true for the ferromagnetic order parameter in subfigure (e), except for $\lambda=0$ where it vanishes everywhere. Maybe somewhat surprisingly the string order parameter is non-zero in both phases but vanishes at the phase transition (and for $\lambda=0$ also in the dimerized phase as well as in the extended critical phase). This can be inferred from subfigure (f).

  Subfigure (g) has been included to illustrate the effect of the deformation on the anisotropy of the spins. In view of $\vec{S}_i^2=2$ one expects $\langle(S_i^z)^2\rangle=\frac{2}{3}$ in the isotropic case. Values above that threshold indicate a preference for the $z$-direction while lower values indicate a preference for the $xy$-plane. We recognize that the $xy$-plane is preferred in the Haldane phase while the $z$-direction is preferred in the dimerized phase. At the phase boundary at $\theta=-\frac{\pi}{4}$ the isotropy surprisingly seems to be restored (on the level of the ground state). When approaching the ferromagnetic region at $\theta\approx-\frac{3\pi}{4}$, the value approaches $1$ which corresponds to Ising variables $S^z=\pm1$. Generally, the degree of anisotropy grows as $\lambda$ increases. All these findings are consistent with our analysis of the crystal limit in Section~\ref{sc:Limit}.
  
  As a sanity check we have also determined the scaling of the entanglement entropy and the correlation length at $\theta=-\frac{\pi}{4}$ as a function of the bond dimension in subfigures (h) and (i). For $\lambda=0$ one clearly recovers the expected relations~\eqref{eq:EntanglementScaling} with $c\approx1.51$, consistent with a $SU(2)_2$ WZW model. For $\lambda=0.5$ there appears to be a small yet systematic deviation from the critical scaling. This deviation becomes even more prominent for $\lambda=0.7$ and $\lambda=1.0$, confirming that the transition is first and not second order.
  
  Our numerical results are inconclusive close to the phase transitions where observables for neighboring points appear to be somewhat uncorrelated. This concerns, especially, the potential critical region for $\theta\lesssim\frac{\pi}{2}$ which, for $\lambda=0$ extends up to $\theta=\frac{\pi}{4}$ but seems (if it still exists) to be significantly reduced for $\lambda>0$. At these points and regions the entanglement of the ground states is likely too large to be accurately captured by iDMRG with a fixed (or at least our chosen) bond dimension.

  Let us finally turn our attention to the question of the existence of SPT phases. Figure~\ref{fig:ESall} shows the conventional and the $q$-deformed symmetry-resolved entanglement spectrum for $\lambda=0.5$ and six values of the parameter $\theta$. These values of $\theta$ correspond to the \qAKLT point and five generic choices (away from any phase transition). The conventional entanglement spectrum shows no signature of systematic degeneracies anywhere in the phase diagram, just a systematic tilt in dependence on the $S^z$-quantum number reminiscent of what is known for higher-spin \qAKLT states \cite{Quella:2020PhRvB.102h1120Q}, see Figure~\ref{fig:ES_tilt}. From the perspective of Ref.~\cite{Pollmann:PhysRevB.81.064439} there is no reason to assume that the system would reside in an SPT phase for any choice of $\theta$.
  
  However, the situation is entirely different for the $q$-deformed entanglement spectrum which aims to correct the systematic tilt in a specific fashion. In that case we clearly see that the tilt is removed giving rise to a degenerate entanglement spectrum, with the exception of truncation effects due to finite bond dimension. To be precise there is an even degeneracy for all investigated choices of $\theta$ in the interval $(-\frac{\pi}{4},\frac{3\pi}{8}]$. Following Ref.~\cite{Quella:2020PhRvB.102h1120Q} we interpret this as a signature of non-trivial topology and we call the associated phase a \qHaldane phase. On the other hand there is no (or just an odd) degeneracy for $\theta=-\frac{5\pi}{8}$ and $\theta=-\frac{3\pi}{8}$ which are located in the dimerized phase. Following Ref.~\cite{Quella:2020PhRvB.102h1120Q} we interpret this as a signature of trivial topology.
  
  The attentive reader might have noticed that the statements about the degeneracy become somewhat ambiguous in the higher parts of the entanglement spectrum. This is the consequence of an interesting subtlety of the numerical simulations that is also likely to render a more detailed analysis of the phase transitions futile using the present method.
  
  Indeed, since the code is not preserving $\qsu$ symmetry it may happen that individual $\qsu$-multiplets are not kept with all their states during singular value decompositions and subsequent truncations that are part of the iDMRG algorithm. This phenomenon also occurs in the undeformed case but it is more pronounced and less random in the presence of the $q$-deformation due to the additional tilt of the ordinary entanglement spectrum, see Figure~\ref{fig:ESall}. This tilt makes it not only more likely that individual states of a single multiplet are removed but moreover the way the multiplets are destroyed is consistently biased towards either positive or negative $S^z$-eigenvalues. We therefore believe that a $\qsu$-covariant iDMRG code is required to analyze the vicinity of phase transitions where the entanglement entropy -- as a function of the bond dimension -- is known to grow beyond all bounds \cite{Pollmann:2009PhRvL.102y5701P}.

\section{One symmetry to rule them all}

  In the final section of this paper we would like to promote an unconventional perspective on SPT phases, specifically those with continuous symmetry. The classification of SPT phases rests on studying the connectivity properties of systems in a suitable space of invariant Hamiltonians \cite{Schuch:1010.3732v3}. Without symmetries this would just be the space of `all' Hamiltonians.\footnote{For physical and mathematical reasons one usually also imposes certain types of locality conditions.} However, depending on the precise type of symmetry this space is further constrained and obviously these constraints change the connectivity properties and hence the classification.
    
  In the case of the $q$-deformed bilinear-biquadratic spin chains discussed in this paper we deal with families of Hamiltonians that have {\em different protecting symmetries} (the quantum groups $\cU_q[su(2)]$ are not isomorphic for different values of $q\geq1$) but still exhibit precisely {\em the same topological features and an identical classification of topological phases}. Since different values of $q\geq0$ are continuously connected this suggests to regard all (a priori distinct) topological phases that are characterized by `the same' topological invariant as one single topological phase with respect to a family of symmetries, all of which have essentially `identical' properties. We wish to call such an extended topological phase a `\qSPT' phase. In practice this means that the topological phase protected by SO(3) symmetry (or any one of its $q$-deformations) is significantly enlarged -- it even increases in dimension. Physically, the possibility of having \qSPT phases is highly relevant since the properties of a given physical system may enjoy {\em strict protection} under a much larger set of deformations than initially anticipated. As we have shown in this paper this is certainly the case for the family of $q$-deformed bilinear-biquadratic spin chains.
  
  The phenomenon just described should also occur for other symmetries that admit deformations that allow for a continuous deformation of essential representation theoretic data such as dimensions of irreducible representations or tensor product decompositions. In particular, it applies to $q$-deformations $\qg$ (with $q>0$) of other simple Lie algebras $\g$ \cite{Quella:2020Draft} (or rather the deformations of the associated Lie groups).\footnote{The situation is quite different for $q$ on the unit circle since the representation theory changes drastically at roots of unity (see, e.g., \cite{Klimyk:MR1492989}).}
  
  It would be interesting to understand whether the notion of a \qSPT phase can also be defined in the context of discrete symmetries or duality-type symmetries. For example, it is known that the (group algebra of the) symmetric group $S_N$ admits a $q$-deformation to a so-called Hecke algebra $H_N(q)$. While this algebraic structure is, perhaps, not so familiar in the condensed matter community it plays an important role in the context of ($q$-deformations) of the Haldane-Shastry model \cite{Bernard:1993JPhA...26.5219B,Lamers:2020arXiv200413210L}.

\section{Conclusions and Outlook}

  In this paper we explored the phase diagram of the anisotropic bilinear-biquadratic spin-$1$ quantum spin chain with $\qsu$ quantum group symmetry. We confirmed that the phase diagram closely resembles that of the undeformed chain except for a significant diminishment or disappearance of the extended critical phase, see Figure~\ref{fig:PhaseDiagram}. We also confirmed the structure of the phase diagram analytically in the extreme anisotropic limit $q\to\infty$, also known as the crystal limit.
  
  While many of our results could be expected, at least qualitatively, it is worth highlighting the existence of a chiral Haldane phase (here called \qHaldane phase) that we could identify unambiguously using a recently predicted two-fold degeneracy in a $q$-deformation of the entanglement spectrum \cite{Quella:2020PhRvB.102h1120Q}. This establishes the $q$-deformed bilinear-biquadratic chain as a prototype of an SPT phase that is not protected by standard symmetries but rather by a generalized symmetry. At the same time, our work is an important proof of evidence that physical systems do not need to be strongly entangled in order to exhibit topological features. Indeed, as the consideration of the regime $q\gg1$ shows, the entanglement may be arbitrarily low as long as a symmetry, in this case $\qsu$, ensures that a certain amount of short-range entanglement remains.
  
  There are many directions and open questions in the context of the $q$-deformed bilinear-biquadratic spin-$1$ chain that deserve further investigation. First of all, it should be possible to gain additional analytical insights from the extreme anisotropic limit by letting the parameter $q$ be large but finite. One could then construct ground states, correlation functions, ($q$-deformed) entanglement spectrum and other quantities perturbatively. This procedure has proved very powerful in the context of quantum integrable models \cite{Davies:1993CMaPh.151...89D} (even though admittedly with an {\em affine} quantum group symmetry).
  
  Another important point is concerning the nature of the phase transitions in Figure~\ref{fig:PhaseDiagram}. Preliminary simulations concerning the scaling of the entanglement entropy as a function of the bond dimension did not result in a clear picture, potentially due to (symmetry-breaking) cut-off effects when performing singular value decompositions and entanglement truncations in the process of iDMRG. The best way to eliminate these effects would be the development of a $\qsu$-invariant DMRG or iDMRG code, in analogy to those that exist for ordinary non-abelian symmetries such as $su(2)$ \cite{McCulloch:2002EL.....57..852M,Weichselbaum:2012AnPhy.327.2972W}.
  
  Besides providing a better picture of what happens in the extended critical region upon introducing the anisotropy such a code would also be able to shed light on the role of the integrable XXZ chain at $\theta=-\frac{\pi}{4}$ which separates the \qHaldane and the dimerized phase. This chain is referred to as being massive in the literature even though its entanglement scaling resembles that of a gapless critical point, more precisely an $SU(2)_2$ WZW model \cite{Weston:2006JSMTE..03L.002W}. The potentially non-differentiable ground state energy in Figure~\ref{fig:PTall} as well as analysis of the limit $\lambda\to\infty$ support a first order quantum phase transition, consistent with the presence of a gap. On the other hand at this particular point in the phase diagram inversion and time-reversal symmetry are restored and the expectation values of $\langle(S_i^\alpha)^2\rangle$ also suggest restoration of isotropy, all necessary conditions for a description in terms of a $SU(2)$ WZW model.

  In another direction of research it would be interesting to consider higher spin representations or spin ladders. In this paper we have focused on the spin-1 case. However, isotropic chains have also been considered for spin-2 for instance \cite{Tu:PhysRevB.78.094404,Tu:PhysRevB.80.014401}. In that case there are two different types of AKLT states in different topological classes, one with spin-$1$ and one with spin-$\frac{3}{2}$ boundary spins, and the phase diagram is very rich. The analysis of $q$-deformations of these systems would be another natural application of the methods presented in this paper.

  From a more conceptual perspective the most urgent issue is to find and explore other generalized symmetries that lead to \qSPT or analogues of \qHaldane phases. It is known that quantum groups of the form $\qg$ \cite{Quella:2020Draft} give rise to such phases, where $\g$ is a simple Lie algebra, but it is currently unclear whether this also holds for multi-parameter deformations or duality-type symmetries where a group transformation is accompanied by a change of coupling as suggested in Ref.~\cite{Quella:2020PhRvB.102h1120Q}.

\begin{acknowledgments}
  The author would like to thank Murray Batchelor and Rafael Nepomechie for useful feedback, Robert Pryor and Caleb Smith for interesting discussions in a closely related context, Johannes Hauschild for answering questions concerning TeNPy's implementation of symmetry-resolved entanglement spectra and Robert Weston for correspondence on the integrable XXZ chain. This research was conducted by the Australian Research Council (ARC) Centre of Excellence for Mathematical and Statistical Frontiers (ACEMS, project number CE140100049) and partially funded by the Australian Government. The author also acknowledges funding from the ARC Discovery Project DP180104027.
\end{acknowledgments}


\def\cprime{$'$}

\end{document}